\documentclass[11pt]{article}

\usepackage[left=2cm, right=2cm, top=2.5cm, bottom=2.5cm]{geometry}
\usepackage[x11names]{xcolor}
\usepackage{fancyhdr, amssymb, cancel, amsmath, graphicx, pgfplots, tikz}

\newcommand{\stylecolor}{black}

\usepackage[colorlinks=true, urlcolor=violet, linkcolor=blue, citecolor=red, hyperindex=true, linktocpage=true]{hyperref}

\usepackage[explicit]{titlesec}

\newcommand*\sectionlabel{}
\titleformat{\section}
  {\gdef\sectionlabel{}
   \Large\bfseries\scshape}
  {\gdef\sectionlabel{\thesection. }}{0pt}
  {\begin{tikzpicture}[remember picture,overlay]
	\draw (-0.2, 0) node[right] {\textsf{\sectionlabel#1}};
	\draw[thick] (0, -0.4) -- (\textwidth, -0.4);
       \end{tikzpicture}
  }
\titlespacing*{\section}{0pt}{15pt}{20pt}

\newcommand*\subsectionlabel{}
\titleformat{\subsection}
  {\gdef\subsectionlabel{}
   \large\bfseries\scshape}
  {\gdef\subsectionlabel{\thesubsection.\ \  }}{0pt}
  {\begin{tikzpicture}[remember picture,overlay]
    	\draw (-0.15, 0) node[right] {\textsf{\subsectionlabel#1}};
       \end{tikzpicture}
  }
\titlespacing*{\subsection}{0pt}{10pt}{10pt}

\newcommand*\subsubsectionlabel{}
\titleformat{\subsubsection}
  {\gdef\subsubsectionlabel{}
   \bfseries\scshape}
  {\gdef\subsubsectionlabel{\thesubsubsection.\ \  }}{0pt}
  {\begin{tikzpicture}[remember picture,overlay]
    	\draw (-0.15, 0) node[right] {\textsf{\subsubsectionlabel#1}};
       \end{tikzpicture}
  }
\titlespacing*{\subsubsection}{0pt}{7pt}{7pt}

\pgfplotsset{every axis legend/.append style={at={(1.02,1)},anchor=north west}}

\newcommand{\titletext}{Memory matrix theory of\\ magnetotransport in strange metals}

\begin{document}

\allowdisplaybreaks

\pagestyle{fancy}
\renewcommand{\headrulewidth}{0pt}
\fancyhead{}

\fancyfoot{}
\fancyfoot[C] {\textsf{\textbf{\thepage}}}

\begin{equation*}
\begin{tikzpicture}
\draw (0.5\textwidth, -3) node[text width = \textwidth] {{\huge \begin{center} \color{\stylecolor} \textsf{\textbf{\titletext}} \end{center}}};
\end{tikzpicture}
\end{equation*}
\begin{equation*}
\begin{tikzpicture}
\draw (0.5\textwidth, 0.1) node[text width=\textwidth] {\large \color{black} $\text{\textsf{Andrew Lucas$^a$ and Subir Sachdev$^{a,b}$}}$};
\draw (0.5\textwidth, -0.5) node[text width=\textwidth] {\small  $^a$\textsf{Department of Physics, Harvard University, Cambridge, MA 02138, USA}};
\draw (0.5\textwidth, -1) node[text width=\textwidth] {\small  $^b$\textsf{Perimeter Institute for Theoretical Physics, Waterloo, Ontario N2L 2Y5, Canada}};
\end{tikzpicture}
\end{equation*}
\begin{equation*}
\begin{tikzpicture}
\draw (0.5\textwidth, -6) node[below, text width=0.8\textwidth] {\small 
We model strange metals as quantum liquids without quasiparticle excitations, but with slow momentum relaxation, and with slow diffusive dynamics of a conserved charge and energy. 
General expressions are obtained for electrical, thermal and thermoelectric transport in the presence of an applied magnetic field using the memory matrix formalism.
In the appropriate limits, our expressions agree with previous hydrodynamic and holographic results.  We discuss the relationship of such results to thermoelectric and Hall transport measurements in the strange metal phase of the hole-doped cuprates. 
};
\end{tikzpicture}
\end{equation*}
\begin{equation*}
\begin{tikzpicture}
\draw (0, -13.1) node[right, text width=0.5\paperwidth] {\texttt{lucas@fas.harvard.edu \\ sachdev@g.harvard.edu}};
\draw (\textwidth, -13.1) node[left] {\textsf{\today}};
\end{tikzpicture}
\end{equation*}

\tableofcontents

\section{Introduction}
\label{sec:intro}

The strange metal phase of the hole-doped cuprate superconductors is the most important realization of quantum matter
not amenable to a quasiparticle description \cite{lt,sk}. Apart from the well-known linear in temperature ($T$) resistivity, 
the strange metal has a Hall angle $\tan (\theta_{\mathrm{H}}) \sim 1/T^2$ \cite{chienexp,pwa}. This combination cannot be reproduced in a Boltzmann 
theory of charge-carrying
quasiparticles with a long lifetime; such a theory yields a $\tan (\theta_{\mathrm{H}})$ inversely proportional to the resistivity, 
and so these observations rule out the transport of charge by any fermionic quasiparticle, and not just those
with the same quantum numbers as the electron.

A general hydrodynamic approach to magnetotransport in strange metals was introduced by Hartnoll {\em et al.} in \cite{hkms}, and compared
to Nernst measurements in the strange metal of the cuprates. 
The main assumption of their theory, apart from the absence of any quasiparticle excitations, was that there was a slow mode associated with
the decay of the total momentum. Such a slow mode is indeed invariably present in proposed field-theoretic models of strange metals \cite{raghu,patel}, 
and the decay arises from perturbations which break the continuous translational symmetry of the field theory (umklapp scattering or impurities).
Combining this momentum mode with reasonable assumptions on the diffusion of the conserved U(1) charge and energy densities, general results were
obtained by Hartnoll {\em et al.} \cite{hkms} for the charge and thermoelectric transport coefficients of a two-dimensional strange metal in the presence of a static magnetic field, $B$. 
Note that we are using the phrase ``charge density'' here to refer to the conserved density of a global U(1) symmetry. In the application to the cuprates, there are
also long-range Coulomb interactions associated with such a conserved density, and we will briefly discuss this in Section~\ref{sec:diffusive}.

Blake and Donos \cite{blakedonos} have recently argued that the magnetotransport framework of Hartnoll {\em et al.}  is compatible with the measurements of the 
longitudinal and Hall conductivities in the cuprates, and also provided a solvable holographic model for magnetotransport in the presence of momentum relaxation.

However, an important concern is that
the present derivations of the above results for strange metals rely on models which are rather
far removed from the microscopic situation in the cuprates. One approach \cite{hkms} begins from a quantum critical point with a relativistic structure, 
and then breaks Lorentz invariance weakly by a chemical potential, the temperature, and the applied magnetic field; the equations were then
derived using a hydrodynamic gradient expansion restricted by the requirements of the positivity of entropy production. The other approach \cite{hkms,blakedonos} uses holographic models of strange metals represented by gravitational theories in a spacetime with an extra dimension, 
whose field theory duals are not well understood.

In the present paper, we will provide a derivation for the equations for magnetotransport in a normal-state metal using the memory matrix approach \cite{zwanzig,mori,forster}.    
Although the memory matrix framework has typically been applied to systems where the slow relaxation of momentum dominates magnetotransport and the diffusion of conserved quantities is negligible, we generalize and study the case where the diffusive dynamics is not negligible and must be consistently included in the memory matrix.    In particular, we will focus on the consequences of the diffusion of exactly conserved heat, as well as an exactly conserved U(1) charge.  Our main results may be found in (\ref{eq:mainres}) and (\ref{eq:mainres2}), though there is a large amount of notation that must be explained, and so we defer their presentation.   
Our assumptions on the momentum mode, and the charge and energy diffusion will be the same as those in \cite{hkms}, but we will not assume any relativistic
structure on the underlying theory.   Using certain assumptions, our results agree with those of \cite{hkms}, and with the recent holographic papers \cite{holot1, holot2, holot3}, and shed light into a  discrepancy between the two.   Furthermore, our results can, in principle, be applied directly to microscopic models appropriate for the cuprates \cite{raghu,patel,DHSS15,patel2}, as we will describe more completely in Section~\ref{sec:conc}.

\subsection{Transport Without Quasiparticles}
\begin{figure}
\begin{center}
\includegraphics[width=130mm]{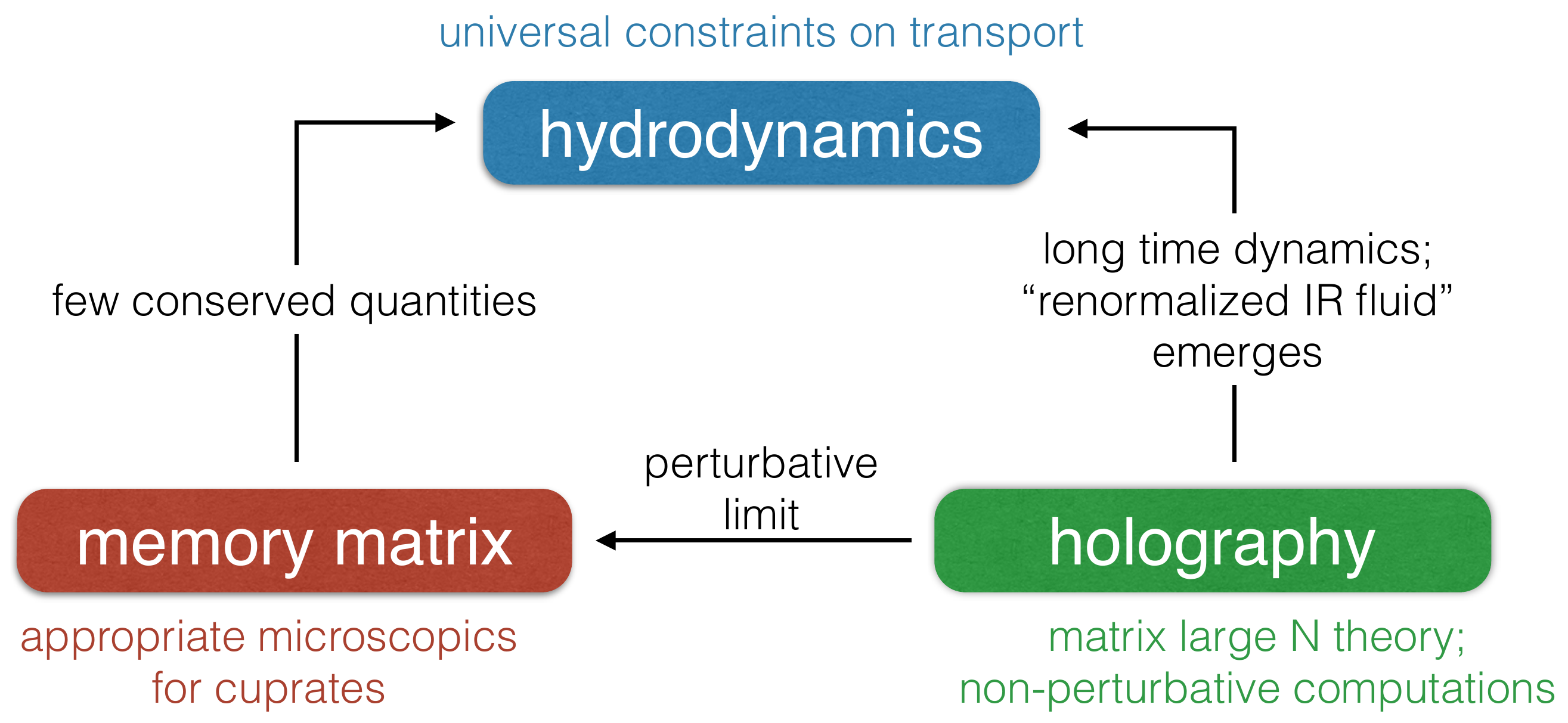}
\end{center}
\caption{An illustration of the connections between hydrodynamics, memory matrices, and holography.}
\label{fig:framework}
\end{figure}

Let us briefly review our current understanding of theoretical frameworks for describing transport in strongly correlated systems without quasiparticles 
in two (or more) spatial dimensions.
An overview is also provided in Fig.~\ref{fig:framework}.

The simplest way of describing transport without reference to quasiparticles is, in fact, a very old framework -- hydrodynamics.    The modern understanding of hydrodynamics is that it describes the long wavelength, long time dynamics of a system close to thermal equilibrium, when there are a small number of conserved quantities \cite{kadanoff}.      We will be liberal, and also allow for some of these conserved quantities to decay on long time scales, while still maintaining the name ``hydrodynamics".    All we insist upon is that this list of conserved quantities is finite; this is in contrast to the typical condensed matter paradigm of a Fermi liquid, where occupation numbers at every single wave number are long lived quantities. For non-Fermi liquids, 
it is believed that generic higher dimensional  theories do not admit infinite families of (nearly) conserved quantities at strong coupling.

 Hydrodynamics proves to be a very powerful \emph{framework} for describing the dynamics of systems without reference to quasiparticles.    However, we must emphasize that hydrodynamics is an \emph{incomplete} description.   It provides a set of constraints that any (to date) reasonable (2+1)-dimensional quantum field theory at finite density and temperature, which is approximately translation invariant, must obey, and a universal framework within which we can interpret a microscopic calculation.    But it does not give us particular values or temperature dependence for any microscopic coefficients.    

So now we need a way of obtaining microscopic coefficients.    A traditional framework for doing this is the memory matrix framework \cite{zwanzig, mori}; see \cite{forster} for a thorough review.   We will review it in this paper, but for now we emphasize that it is in principle an exact, microscopic calculation.  Its usefulness is that it can be efficiently approximated in a hydrodynamic regime where there are only a few quantities which do not quickly relax to thermal equilibrium.   

The main purpose of this paper is to clarify and sharpen the connections between the memory matrix framework for transport, and hydrodynamic descriptions of transport.  In particular, we will not need to add a phenomenological momentum relaxation time in hydrodynamics, and compute this coefficient separately using the memory matrix formalism.    We will also point out the microscopic computations which  allow us to compute all phenomenological coefficients within hydrodynamics.

There has also been, in recent years, a third framework which allows for transport computations:   gauge-gravity duality.   This allows one to study a strongly coupled ``large $N$ matrix" quantum field theory, directly at finite temperature and density, in real time, by studying a classical gravity theory in one higher dimension, in particular black hole backgrounds.  Ref.~\cite{lucasproof} has recently shown the equivalence between holographic and memory function calculations in zero magnetic field, in the regime where momentum relaxation is slow (this provides the bottom arrow in Fig. \ref{fig:framework}).  One advantage of holographic methods is that the calculations are formally valid beyond the regimes of validity of hydrodynamics or memory functions.  However, we point out in this paper that a wide variety of holographic results for transport, computed in regimes where hydrodynamics need not be valid, can nonetheless be understood within the framework we derive.

\subsection{Outline of the Paper}

We will begin in Section~\ref{sec:hydro} by recalling the main hydrodynamic results of \cite{hkms} for electrical transport in a simple manner, and using  notation suitable
for our memory matrix approach. The basic memory matrix formalism will be introduced in Section~\ref{sec:formalism}, along with its
physical interpretation. Section~\ref{sec:mhd} contains our main new results on magnetotransport in the memory matrix formalism under the simplifying assumption that the diffusion of heat has decoupled from the diffusion of charge and relaxation of momentum:  this simplification will help us to elucidate many of the subtleties that arise in the memory matrix framework when describing magnetotransport.  Section \ref{appa} relaxes the assumption that heat diffusion decouples, and Section \ref{sec:therm} describes thermal and thermoelectric transport.

\section{Hydrodynamics}
\label{sec:hydro}

Let us begin by computing electrical transport in a generic quantum field theory without quasiparticles, describable by hydrodynamics,
following the approach of \cite{hkms}.   More precisely, we focus on a theory with exactly conserved energy and U(1) charge, and approximately conserved momentum, where the only dynamics on long time scales are momentum relaxation on the time scale $\tau$, and charge/energy diffusion on the time scales $\lambda^2/D$, with $\lambda$ the wavelength of the fluctuation and   $D$ an eigenvalue of the generalized diffusion matrix coupling heat and charge diffusion.\footnote{Note that with momentum relaxation, there is no longer a propagating sound mode as $k\rightarrow 0$, so indeed the dynamics of charge and energy become governed by diffusion.}

We wish to compute the electrical conductivity matrix, defined as \begin{equation}
J_i = \sigma_{ij}E_j.
\end{equation}We assume that we have an isotropic, parity-symmetric theory in two spatial dimensions, which constrains\begin{subequations}\begin{align}
\sigma_{xx} &= \sigma_{yy}, \\
\sigma_{xy} &= -\sigma_{yx}.
\end{align}\end{subequations}

Momentum conservation within a hydrodynamic framework -- accounting for momentum relaxation -- gives us the equation \begin{equation}
\partial_t \Pi_x + \partial_x P = -\frac{\Pi_x}{\tau} + \mathcal{Q}E_x + B\epsilon_{ij}J^j
\end{equation} 
where $\mathcal{Q}$ is the background ``charge'' density, $\tau$ is the momentum relaxation time, $B$ is the external magnetic field, $P$ is the pressure and $\Pi_x$ is the $x$-momentum density; a precise definition of $\mathcal{Q}$, applicable to lattice models, appears later in (\ref{defQ}). 
We have assumed that the velocities relevant for hydrodynamics are small and may be treated non-relativistically.   As we are interested in perturbing the system with a spatially homogeneous electric field, oscillating at frequency $\omega$, momentum conservation simplifies to \begin{equation}
\left(-\mathrm{i}\omega + \frac{1}{\tau}\right)\mathcal{M}\delta v_i = \mathcal{Q} \delta E_i + B \epsilon_{ij} \delta J_j   \label{eqv}
\end{equation}where $\mathcal{M}$ is the analogue of ``mass density" (namely, the momentum density at small velocities is $\mathcal{M}\delta v_i$).    The memory matrix definition of $\mathcal{M}$ is in (\ref{defM}).
 The momentum relaxation time, $\tau$, is -- like all thermodynamic coefficients -- undetermined within hydrodynamics.   Unlike thermodynamic coefficients like $\mathcal{M}$ and $\mathcal{Q}$, however, $\tau$ is extrinsic and sensitive to the precise mechanism by which momentum relaxes, as we will see explicitly within the memory matrix formalism.

(\ref{eqv}) gives us an equation relating $\delta v_i$, $\delta J_i$ and $\delta E_i$ -- our perturbatively small quantities within linear response.    $\delta J_i$ may be expressed in terms of $\delta v_i$ and $\delta E_i$, in an isotropic and parity-symmetric (up to magnetic fields) theory, as \begin{equation}
\delta J_i = \mathcal{Q}\delta v_i + \sigma_{\textsc{q}} \left(\delta E_i + B \epsilon_{ij} \delta v_j\right).   \label{eqj}
\end{equation}  
Here $\sigma_{\textsc{q}}$ is an intrinsic ``quantum critical'' conductivity which measures charge transport independent of the momentum mode; a precise definition relating it to the charge diffusivity appears later.   
Combining (\ref{eqv}) and (\ref{eqj}) we may relate $\delta J_i$ to $\delta E_i$, and we find 
\begin{subequations}\begin{align}
\sigma_{xx} &= \frac{(\tau^{-1}-\mathrm{i}\omega)\mathcal{M}\sigma_{\textsc{q}} + \mathcal{Q}^2 + B^2\sigma_{\textsc{q}}^2}{\mathcal{Q}^2B^2 + ((\tau^{-1}-\mathrm{i}\omega)\mathcal{M}+B^2\sigma_{\textsc{q}})^2} \mathcal{M}\left(\frac{1}{\tau}-\mathrm{i}\omega\right), \\
\sigma_{xy} &=  \frac{2(\tau^{-1}-\mathrm{i}\omega)\mathcal{M}\sigma_{\textsc{q}} + \mathcal{Q}^2 + B^2\sigma_{\textsc{q}}^2}{\mathcal{Q}^2B^2 + ((\tau^{-1}-\mathrm{i}\omega)\mathcal{M}+B^2\sigma_{\textsc{q}})^2} B\mathcal{Q}.
\end{align}
\label{eq:mainres}
\end{subequations}
The answer simplifies when $B=0$ to the form $\sigma_{xy}=0$ (by parity symmetry) and \begin{equation}
\sigma_{xx} = \sigma_{\textsc{q}} + \frac{\mathcal{Q}^2\tau}{\mathcal{M}(1-\mathrm{i}\omega \tau)}.  \label{eqb0}
\end{equation}

Though in a system without Galilean invariance one might be concerned that the $\mathcal{Q}$s in (\ref{eqv}) and (\ref{eqj}) need not be the same, we will carefully define $\mathcal{Q}$ via the memory matrix framework and show that (\ref{eq:mainres}) is indeed (at leading order) correct.

These results were first derived in \cite{hkms}, for the case of a quantum critical theory, and were argued in \cite{blakedonos} to be compatible
with Hall angle measurements in the strange metal of the cuprates.   In fact, the structure here is identical, and is valid for any isotropic theory, whether or not it is Lorentz invariant.    We even need not assume that diffusive charge and heat fluctuations are decoupled (as would be the case when the underlying theory has charge conjugation symmetry) -- heat transport could only alter the equations above if temperature gradients were non-vanishing, which (by assumption) does not happen.   It is a simple task to compute additional thermal and thermoelectric coefficients, but we will not do so using hydrodynamics. 

We remarked previously that the dynamics which governs $\sigma_{ij}$ is the only dynamics in the system occuring on long time scales.   However, the expressions for $\sigma_{ij}$ are rather complex!   This is because there are four slow time scales in the problem: the slow driving time scale $1/\omega$, and the three time scales associated with the specific quantum field theory, and its realization: $\tau$, $1/\omega_{\mathrm{c}}$, and $1/\omega_{\textsc{q}}$, where $\omega_{\mathrm{c}}$ is the cyclotron frequency \begin{equation}
\omega_{\mathrm{c}} = \frac{\mathcal{Q}B}{\mathcal{M}},
\end{equation}and $\omega_{\textsc{q}}$ is a ``quantum frequency" given by \begin{equation}
\omega_{\textsc{q}} = \frac{\mathcal{Q}^2}{\mathcal{M}\sigma_{\textsc{q}}}.
\end{equation}
Within hydrodynamics, and the memory function formalism, we  only work to leading order in $\omega$ -- but to all orders in the dimensionless numbers $\omega\tau$, $\omega/\omega_{\mathrm{c}}$, $\omega/\omega_{\textsc{q}}$.

As $\tau^{-1}$ scales with the strength of the coupling to the operator which breaks translational symmetry (as we will explicitly see later), and $\omega_{\mathrm{c}}\sim B$, it is evident in what physical limit these two quantities would be perturbatively small.    It is less clear how $\omega_{\textsc{q}}$ would generically be perturbatively small.    One mechanism is that the densities of the theory ($\mathcal{Q}$ and $\mathcal{M}$) are small (with $\mathcal{Q}/\mathcal{M}$, the strong-coupling analogoue of the fixed charge-to-mass ratio of the electron, fixed).    Another is that $\sigma_{\textsc{q}}$ is anomalously large -- though this should not be a consequence of weak coupling (i.e., long-lived quasiparticles), for our framework to be valid.

 We should also comment that recently, this result has been obtained holographically at $\omega=0$ \cite{blakedonos} (though it is not expressed in terms of thermo/hydrodynamic quantities) in a set-up which is formally valid even when $\tau$ is comparable to $\tau_{\mathrm{m}}$ -- the time scale associated with microscopic dynamics, and the breakdown of hydrodynamics.  The limiting case (\ref{eqb0}), also at $\omega=0$, is found in various holographic set-ups in \cite{blake1, gouteraux, LS2}.   It is an interesting question whether or not this is a pathological feature of the holographic mechanism of strong momentum relaxation, or a signature that the ``hydrodynamic" regime of strongly coupled transport persists far beyond our first expectations.

\section{Memory Function Formalism}
\label{sec:formalism}

Let us now briefly review the memory function framework.  As we emphasized in the introduction, the method works best when quasiparticles are not long-lived, and the only conserved (or approximately conserved) quantities are charge, energy and momentum.   In recent years this has become a method of choice for studying transport in condensed matter systems without quasiparticles \cite{hkms, rosch, raghu, patel}, including using gauge-gravity duality \cite{hartnollimpure, hartnollhofman, blake2, dsz, lss}.    We work in units with $\hbar=1$.

   Let us consider the set of operators $A,B,C,\ldots$ in a time-translation invariant theory, and correlation functions of the form \begin{equation}
\mathcal{C}_{AB}(t-t^\prime) = (A(t)|B(0)) \equiv T\int\limits_0^{1/T}\mathrm{d}\lambda\; \left\langle A(t)^\dagger B(\mathrm{i}\lambda)\right\rangle  \label{innprod}
\end{equation}with averages over thermal and quantum fluctuations denoted in $\langle \cdots\rangle$.   The Laplace transform of this expression can be shown to be \begin{equation}
\mathcal{C}_{AB}(z) = \frac{T}{\mathrm{i}z}\left[G^{\mathrm{R}}_{AB}(z) - G^{\mathrm{R}}_{AB}(\mathrm{i}0)\right].  \label{cabz}
\end{equation}The retarded Green's function (in real space) is defined as: \begin{equation}
G^{\mathrm{R}}_{AB}(\mathbf{x},t) \equiv \mathrm{i}\Theta(t) \langle  [A(\mathbf{x},t),B(\mathbf{0},0)]\rangle
\end{equation}
where $\Theta$ is the Heaviside step function.  The momentum dependence of the Green's functions have been suppressed in (\ref{cabz}) as we will only be interested in the conductivities, evaluated at zero momentum.  As standard, we take $z$ to lie in the upper half of the complex plane.  We are only computing thermoelectric transport coefficients, for which $G^{\mathrm{R}}_{AB}(\omega \rightarrow 0) \sim \mathrm{i}\omega \sigma_{AB}$, with $\sigma_{AB}$ strictly finite;  thus for us, $G^{\mathrm{R}}_{AB}(\mathrm{i}0)=0$ and can be neglected.   Up to the overall factor of temperature, we recognize $\mathcal{C}_{AB}(z)$ in (\ref{cabz}) as a generalized conductivity between the operators $A$ and $B$, $\sigma_{AB}$.   Some formal manipulations on a Hilbert space of operators using the inner product (\ref{innprod}) give  \begin{equation}
\sigma_{AB}(z) \equiv \frac{1}{T}\mathcal{C}_{AB}(z) = \chi_{AC}[M(z)+N-\mathrm{i}z \chi]^{-1}_{CD}\chi_{DB},  \label{mmf}
\end{equation}where $\chi_{AB}$ is the static susceptibility between the operators $A$ and $B$:\footnote{Differing conventions appear in the literature \cite{kovtun}.   This definition of the (overall sign of the) retarded Green's function is much more standard \cite{hartnollhofman, dsz} and is the common choice in condensed matter physics.}
\begin{equation}
\chi_{AB} \equiv G^{\mathrm{R}}_{AB}(\omega=0) = \frac{1}{T} (A|B),   \label{chiab}
\end{equation}  the memory matrix $M_{AB}$ is defined as \begin{equation}
M_{AB}(z) = \frac{\mathrm{i}}{T}\left(\dot{A}\left| \mathfrak{q}(z-\mathfrak{q}L\mathfrak{q})^{-1}\mathfrak{q}\left|\dot{B}\right)\right.\right.,  \label{memmatrix}
\end{equation}$L = [H,\circ]$ is the Liouvillian operator with $H$ the Hamiltonian, $\mathfrak{q}$ is the projection operator \begin{equation}
\mathfrak{q} = 1-\frac{1}{T}\sum_{AB} \chi_{AB}^{-1} |A)(B|,
\end{equation}and\begin{equation}
N_{AB} \equiv \chi_{A\dot{B}} = - \chi_{\dot{A}B}.
\end{equation}Note that $N_{AB}$ vanishes identically in a time-reversal invariant theory.

We will show that for the magnetotransport problem of interest in this paper, the memory matrix effectively truncates to a $6\times 6$ matrix, given as $\mathfrak{m} = M+N-\mathrm{i}\omega\chi$ in (\ref{mmfinal}).

\subsection{Hydrodynamic Interpretation}
The usefulness of the memory function formalism  arises when the infinite dimensional matrix $M$ has a finite number of  parametrically small eigenvalues.   This allows us to truncate the matrix to a small, finite dimensional object.

To give some physical intuition into (\ref{mmf}), let us consider the dynamics of long-lived conserved quantities from the viewpoint of hydrodynamics.   Let $X_A$ denote quantities which are conserved on very long time scales, and let $U_A$ be the conjugate thermodynamic variable.   For simplicity, we define $X_A$ by shifting $X_A$ by a constant so that in equilibrium, $\langle X_A\rangle = 0$.  In thermodynamics, we can then relate \begin{equation}
X_A = \chi_{AB}U_B,
\end{equation}where $\chi$ is the same static susceptibility matrix as before, and we drop expectation values within hydrodynamics.    The transport equations read \begin{equation}
\dot{X}_A = -M_{AB}U_B -N_{AB}U_B + F_A,  \label{dxadt}
\end{equation}where $N_{AB}$, as before, is the static susceptibility between $\dot{A}$ and $B$, $M_{AB}$ is a (for now arbitrary, but symmetric) matrix which denotes the weak relaxation of $X_A$ into an external bath, and $F_A$ is an external driving force.    Now, we relate $F_A$ to the ``thermodynamic field" $E_A$ which drives $X_A$ by \begin{equation}
F_A = \chi_{AB}E_B.
\end{equation}
$E_A$ should be interpreted as  ``the net $\dot{U}_A$" imposed on the system.   Combining these equations together at frequency $\omega$, we obtain \begin{equation}
X_A = \sigma_{AB}E_B
\end{equation}where $\sigma_{AB}$ is given by (\ref{mmf}). 

One aspect of the memory function framework  that  is now much clearer is the physical role of the memory matrix.   Defining the memory matrix as \begin{equation}
M_{AB} \equiv \tau^{-1}_{AC} \chi_{CB},
\end{equation}in an undriven, time reversal invariant theory (where $N=0$), we find \begin{equation}
\dot{X}_A = -\tau^{-1}_{AB} X_B.
\end{equation}Evidently, the memory matrix encodes relaxation times for the average values of the operators $A$ of interest.   It is now evident why we choose the operators $A,B,\ldots$ to be almost conserved quantities.   In this case, the matrix $\tau^{-1}$ will have parametrically small eigenvalues, and as the memory matrix enters the conductivity via an inverse, these quantities will dominate the conductivity.

This derivation is rather abstract, so let us give a simple example to clarify the thermodynamic structure of these equations: the dynamics of a point particle of mass $m$ on a one dimensional line, subject to weak friction, but no potential energy.   In this case, the momentum $p$ is long lived.   The thermodynamic conjugate variable is the velocity $v$, and so the susceptibility is simply the mass.    $N_{pp}=0$, as there is no generic proportionality coefficient between $p$ and $\dot{v}$, and so we are left with $\dot{p} = F - M_{pp}v$.   Of course if $M_{pp}=0$ we simply have Newton's Law, $F = \dot{p} = m\dot{v}$; otherwise, $M_{pp}\ne 0$ provides a frictional force which dissipates momentum into an external bath.  

It is also important to understand the role played by $F$ in this equation.  To compute the ``momentum conductivity" $\sigma_{pp}$ of this particle, we can interpret the external driving by the force $F$ as actually external ``shaking" of the line with acceleration $a_{\mathrm{ext}}=F/m$.  The conductivity relates the response of $p$ to an external source of the time derivative of its conjugate variable.  We find at zero frequency: \begin{equation}
\frac{p}{a_{\mathrm{ext}}} = \frac{mv}{a_{\mathrm{ext}}} = \frac{mF}{M_{pp}a_{\mathrm{ext}}} = \frac{m^2}{M_{pp}},
\end{equation}in accordance with (\ref{mmf}).

Of course, we are probably more familiar with electrical conductivity.   In this case, the thermodynamic conjugate variable to $\mathbf{J}$ is a gauge potential $-\mathbf{A}$.   And as $-\mathrm{d}\mathbf{A}/\mathrm{d}t = \mathbf{E}$,  the electrical conductivity $\sigma_{\mathbf{JJ}}$ relates the response of $\mathbf{J}$ to an external electric field $\mathbf{E}$.   In this case, as we will see in the next section, the analogue of (\ref{dxadt}) become the hydrodynamic equations which describe the dynamics of charge and momentum, which are a bit more cumbersome.

The usefulness of the memory function formalism beyond the hydrodynamic reasoning described above is that we now have explicit expressions for the matrices $M$, $N$ and $\chi$.   In particular, this will allow us to explicitly compute $\tau$, the momentum relaxation time.   We will describe more in the following section what the components of this matrix actually look like.

\section{Dynamics of Charge and Momentum}
\label{sec:mhd}

Let us now discuss how the memory function framework produces the magnetohydrodynamic electrical transport described earlier.   For simplicity, we assume that charge and heat diffusion are decoupled processes at zero charge density -- we relax this assumption in Section \ref{appa}.    There are two slow time scales in the problem:  the time scale over which momentum relaxes $t_{\mathrm{mom}} \sim \tau$, and the time scale over which charge fluctuations diffuse away, $t_{\mathrm{diff}} \sim \lambda^2/D$, with $\lambda$ the wavelength of the fluctuation; on the longest length scales, $t_{\mathrm{diff}}\rightarrow \infty$.   Note that at finite charge density, there will be heat diffusion;  however, the slow diffusion of heat will not be linearly independent from the dynamics of charge or momentum.   Our goal is to compute the electrical conductivity matrix $\sigma_{ij}$ at zero momentum, and small but finite frequency $\omega$.   There are four operators associated with the long lived quantities above which have spin 1 under the SO(2) isotropy of the system:  the momentum vector $\mathbf{P}$, and charge density fluctuations $\nabla n$.    This suggests that our memory matrix should be a $4\times 4$ matrix, as we should only keep track of the (zero wave vector components) of these operators.   Using the notation of the previous section, our goal is to compute the conductivities $\sigma_{J_xJ_x} \equiv \sigma_{xx}$  and $\sigma_{J_xJ_y} \equiv \sigma_{xy}$.  

\subsection{Diffusive Transport}
\label{sec:diffusive}

Let us begin with the case when the charge density $\mathcal{Q}=0$, and the magnetic field $B=0$.   As $B=0$, the time reversal symmetry breaking matrix $N$ will vanish.   In this case, hydrodynamics tells us that the conductivity is governed solely by diffusive processes.  Let us see how this arises in the memory function framework.   In this case, using the isotropy of the system, in order to compute $\sigma_{xx}$, we need only keep track of operators with long time dynamics which are spin 1, and oriented in the $x$ direction:   these are $\partial_x n$ and $P_x$ -- both evaluated at zero wave vector.

To get started, let us compute the susceptibility matrices. Formally, we introduce a velocity $v$, as a source term conjugate to the conserved momentum, so that Hamiltonian maps as
\begin{equation}
H_0 \rightarrow H_0 -  v_x P_x,   \label{h0v}
\end{equation}
where $H_0$ is translationally invariant, and thus commutes with $P_x$, the total momentum in the $x$-direction (we will add terms responsible for momentum decay later).
We may compute the susceptibilities $\chi_{AP_x}$ by studying $\langle A\rangle_{v_x}$ -- the response of the average value of $A$ to a small external source of velocity \cite{kadanoff}:   \begin{equation}
\chi_{P_xP_x} = \frac{\langle P_x\rangle_{v_x}}{v_x V} \equiv \mathcal{M}, \label{defM}
\end{equation}
where $V$ is the spatial volume of the sample.   This formally defines the value of $\mathcal{M}$;  alternatively, $\mathcal{M}$ is related to $G^{\mathrm{R}}_{P_xP_x}$ via (\ref{chiab}).   Note that in the earlier relativistic formalism, $\mathcal{M}$ was equal to the sum of the energy
density and pressure, up to coefficients proportional to the (effective) speed of light.
Analogously, \begin{equation}
\chi_{P_xJ_x} = \frac{\langle J_x\rangle_{v_x}}{v_xV} \equiv \mathcal{Q}, \label{defQ}
\end{equation}
defines the charge density $\mathcal{Q}$.
As the perturbation (\ref{h0v}) does not break translation invariance in $n(\mathbf{x})$: \begin{equation}
\chi_{P_x\partial_x n} = 0.
\end{equation}
Identical results hold for the $yy$ susceptibility matrices.
By spatial parity symmetry, when $B=0$, we find \begin{equation}
\chi_{P_xP_y} = \chi_{\partial_x n \partial_y n} = \chi_{P_x\partial_y n}  = \chi_{P_y\partial_x n} = 0.
\end{equation}

In fact, we now see there is a great simplification when $\mathcal{Q}=0$, because the $\mathbf{P}$ indices of the memory matrix can play no role in the computation of $\sigma_{xx}$, as $\chi_{P_xJ_x}=0$.   This is a bit fast; we will see explicitly later that the necessary components of the memory matrix $M_{P_xA}$ do not vanish fast enough as $\mathcal{Q}\rightarrow 0$ to spoil this argument.

Next, let us compute $\chi_{J_x \partial_x n}$.   The simplest way to proceed here is to choose a finite momentum regulator, and then show that the transport coefficients are independent of the choice of regulator, $k_x$.    We choose $n(x) =  -n(k_x)\sin(k_xx)$,\footnote{The choice of sine waves instead of plane waves keeps things manifestly real, and so is a bit simpler.} and this implies that the overlap with the current operator $J_x(k_x)\cos(k_xx)$, is given by \begin{equation}
\langle \mathbf{J} \rangle = -\sigma_{\textsc{q}} \nabla \mu,  \label{jmu}
\end{equation}as the conjugate field to $n$ is the chemical potential $\mu$ (in the presence of long-range Coulomb interactions, $\mu$ should be
replaced by the electrochemical potential \cite{pines,godby}).   
We conclude\begin{equation}
\chi_{J_x \partial_x n} = \sigma_{\textsc{q}} k_x.
\end{equation}
Crudely speaking, we think of fixing $x$ and sending $k_x\rightarrow 0$ -- one then finds that $n(x) \rightarrow -n(k_x)k_xx$ and $J_x \rightarrow J_x(k_x)$, so that we have a linear density fluctuation sourcing a constant, diffusive current.   But it is easiest to compute the memory matrix with the regulator explicitly finite.

It is a standard result that \cite{kadanoff} \begin{equation}
G^{\mathrm{R}}_{nn}(\mathbf{k},\omega) = \frac{\sigma_{\textsc{q}}k^2}{Dk^2-\mathrm{i}\omega}, \label{Gdiff}
\end{equation}for any quantum field theory which has a hydrodynamic limit, where $n$ will obey a simple diffusion equation at long wavelengths with diffusion constant $D$: $\partial_t n = D\nabla^2 n$ (recall we have assumed charge and heat fluctuations decouple at $\mathcal{Q}=B=0$);
for the case with long-range Coulomb interactions, (\ref{Gdiff}) refers to the irreducible density correlator {\em i.e.\/} the polarizability \cite{pines,godby}.  
We may now use (\ref{mmf}) to obtain $M_{\partial_x n \partial_x n}$.   By time reversal symmetry when $B=0$, $(\partial_x \dot{n}|\partial_x n) = (\partial_x \dot{n}|P_x) = 0$;  thus we take the abstract projection operator $\mathfrak{q}=1$ in (\ref{mmf}).   Using that \cite{forster}\begin{equation}
M_{AB}(z) = \frac{1}{\pi \mathrm{i}}\int\limits_{-\infty}^\infty \mathrm{d}\omega \; \frac{\mathrm{Im}\left(G^{\mathrm{R}}_{\dot{A}\dot{B}}(\omega)\right)}{\omega (\omega-z)},
\end{equation}
and regulating this integral,  we obtain \begin{equation}
M_{\partial_x n \partial_x n}(\omega) = \frac{\sigma_{\textsc{q}}Dk_x^4}{Dk_x^2-\mathrm{i}\omega} \approx \sigma_{\textsc{q}}k_x^2 + \mathrm{i}\omega \chi_{nn} + \cdots,
\end{equation}where we have expanded to lowest non-trivial order in $\omega$, and \begin{equation}
\chi_{nn} = \frac{\sigma_{\textsc{q}}}{D} =  \lim_{k\rightarrow 0} \lim_{\omega\rightarrow 0} \mathrm{Re}\left[G^{\mathrm{R}}_{nn}(\mathbf{k},\omega)\right]
\end{equation}is the static density-density susceptibility.  The memory matrix formalism now gives \begin{equation}
\sigma_{xx} = \frac{\chi^2_{J_x \partial_x n}}{M_{\partial_x n \partial_x n} - \mathrm{i}\omega \chi_{nn}} = \frac{\sigma_{\textsc{q}}^2k_x^2}{\sigma_{\textsc{q}}k_x^2} = \sigma_{\textsc{q}} + \mathcal{O}\left(\omega^2\right).
\end{equation}
This is exactly what we should expect -- the diffusive hydrodynamic coefficient $\sigma_{\textsc{q}}$ is well-known to be the electrical conductivity at $\mathcal{Q}=B=0$.   And importantly, the regulator $k_x$ has decoupled from the answer.   We may thus safely take the limit $k_x\rightarrow 0$.   By spatial isotropy, $\chi_{J_x\partial_x n} = \chi_{J_y\partial_y n}$, etc., and $\chi_{J_x\partial_y n} = 0$, etc.

\subsection{Momentum Relaxation}

Next, let us consider the case where $\mathcal{Q}\ne 0$, and $B=0$.    We have already computed all relevant static susceptibilities, so only the memory matrix $M$ remains.   Spatial isotropy implies that $M_{P_xP_y} = M_{P_x\partial_y n} = 0$, etc., and so we only need to compute $M_{\partial_x n P_x}$ and $M_{P_xP_x}$.      Note that the diffusive form of $M_{\partial_x n \partial_x n}$ and $\chi_{J_x\partial_xn}$ from the previous subsection remains valid even with $\mathcal{Q}\ne 0$, as momentum relaxation kills sound propagation at the longest wavelengths, and charge is still conserved, though the value of $\sigma_{\textsc{q}}$ may change.

We begin with $M_{P_xP_x}$.    Let us consider for simplicity the case where the Hamiltonian of the system is given in $d$ spatial dimensions 
by \begin{equation}
H = H_0 - \int\mathrm{d}^d\mathbf{x}\; h(\mathbf{x}) \mathcal{O}(\mathbf{x}), \label{Pdot}
\end{equation}where $H_0$ is a translationally invariant Hamiltonian, and $\mathcal{O}(\mathbf{x})$ is an arbitrary operator in the quantum field theory.   $h(\mathbf{x})$ is an arbitrary function which should depend on space, and be non-trivial in (almost) all of the plane.   We assume that $h(\mathbf{x})$ is small, so that translational symmetry breaking is weak -- we will see that this is equivalent to the momentum relaxation time $\tau$ being large.    The operator $\dot{P}_i$ is easily computed to be \begin{equation}
\dot{P}_x = \mathrm{i}[H,P_x] = - \int \mathrm{d}^d\mathbf{x} \; h(\mathbf{x}) (\partial_i \mathcal{O})(\mathbf{x}) \label{Pdot2}
\end{equation}
The generalization to the case where multiple operators couple to $\mathbf{x}$-dependent fields is straightforward.   We find \begin{align}
M_{P_xP_x}(z) &= \frac{1}{\pi\mathrm{i}}\int\limits_{-\infty}^\infty \mathrm{d}\omega \; \frac{\mathrm{Im}\left(G^{\mathrm{R}}_{\dot{P}_x\dot{P}_x}(z)\right)}{\omega(\omega-z)}  \notag \\
&= \frac{1}{\pi\mathrm{i}}\int\limits_{-\infty}^\infty \mathrm{d}\omega \; \int \mathrm{d}^d\mathbf{k}_1\mathrm{d}^d\mathbf{k}_2 \; \frac{\mathrm{Im}\left(G^{\mathrm{R}}_{\mathcal{OO}}(\mathbf{k}_1,\mathbf{k}_2,\omega)\right) }{\omega(\omega-z)}  h(\mathbf{k}_1)h(\mathbf{k}_2)  k_{1x}k_{2x}.
\end{align}
As before, we have set $\mathfrak{q}=1$ in this equation:  we assume the operator $\mathcal{O}$ is independent of $\partial_x n$ or $P_x$.    As $h$ is a small parameter, we may safely take $G^{\mathrm{R}}_{\mathcal{OO}}$ to be the Green's functions associated with $H_0$;  translation invariance then implies that it vanishes unless $\mathbf{k}_1+\mathbf{k}_2=\mathbf{0}$.   One finds \cite{hkms,hartnollhofman,lss, lucasproof} \begin{equation}
M_{P_xP_x}(0) \equiv \frac{\mathcal{M}}{\tau} = \lim_{\omega\rightarrow 0} \int \mathrm{d}^d\mathbf{q} \; |h(\mathbf{q})|^2 q_x^2 \frac{\mathrm{Im}\left(G^{\mathrm{R}}_{\mathcal{OO}}(\mathbf{q},\omega)\right)}{\omega},
\end{equation}up to a re-scaling of $h(\mathbf{q})$ to absorb a factor of the spatial volume \cite{lucasproof}.
We will shortly see that the $\tau$ defined here is equal to the $\tau$ defined within hydrodynamics;   with the memory function framework, however, we now have an explicit expression for $\tau$ in terms of the fields $h$.    A similar formula holds for $M_{P_yP_y}$.   For simplicity, we will assume that the system is isotropic and so $M_{P_xP_x} = M_{P_yP_y}$.

We have focused on the $\omega=0$ limit of $M_{P_xP_x}(\omega)$.   This can be justified on very general grounds.    Suppose that $\tau_{\mathrm{m}}$ is the microscopic time scale associated with the quantum dynamics of the system -- for example, in many quantum critical theories, we have $\tau_{\mathrm{m}} = 1/T$.    The Green's function $G^{\mathrm{R}}_{\mathcal{OO}}$ is associated with the quantum dynamics of the Hamiltonian $H_0$, and thus we obtain \begin{equation}
G^{\mathrm{R}}_{\mathcal{OO}}(\omega)= A^{(0)}_{\mathcal{OO}}+ (\omega \tau_{\mathrm{m}}) A^{(1)}_{\mathcal{OO}} +  (\omega \tau_{\mathrm{m}})^2 A^{(2)}_{\mathcal{OO}} + \cdots
\end{equation}
where, for a generic quantum field theory, the coefficients $A^{(1)}_{\mathcal{OO}}$, etc., have no anomalously large coefficients, relative to any others.    This implies that \begin{equation}
M_{P_xP_x}(\omega) = \frac{\mathcal{M}}{\tau}\left[1+\mathcal{C}\omega\tau_{\mathrm{m}} + \cdots\right],
\end{equation}with $\mathcal{C}$ an $\mathcal{O}(1)$ constant, and thus the finite frequency corrections to $M_{P_xP_x}$ are higher order in perturbation theory.   At  $\omega \tau_{\mathrm{m}} \sim 1$, hydrodynamics and the memory function formalism (when truncated to a finite set of operators) both cease to be good approximations for generic theories.

Next, let us discuss $M_{P_x\partial_x n}$.    In the case where the operator $\mathcal{O}$ is charge conjugation symmetric, this matrix element vanishes by charge conjugation symmetry.   We discuss the more general case in Appendix \ref{appb}.  

Putting everything together, we obtain $\sigma_{xy}=0$ (by parity symmetry) and, since the memory matrix is diagonal up to this point: \begin{equation}
\sigma_{xx} = \frac{\chi_{J_x\partial_x n}^2}{M_{\partial_x n \partial_x n} - \mathrm{i}\omega \chi_{nn}} + \frac{\chi_{J_xP_x}^2}{M_{P_xP_x} - \mathrm{i}\omega \chi_{P_xP_x}} = \sigma_{\textsc{q}} + \frac{\mathcal{Q}^2\tau}{\mathcal{M}(1-\mathrm{i}\omega \tau)}
\label{ss1}
\end{equation}
In a theory without momentum relaxation, we have $\tau \rightarrow \infty$; then the real part of the second term in (\ref{ss1}) is proportional to $\delta (\omega)$. In such a theory we define $\sigma_{\textsc{q}}$ to be equal to 
$\sigma_{xx}$ minus the delta function contribution, and this definition can be used
to compute $\sigma_{\textsc{q}}$ in a particular model of a strange metal.

\subsection{Magnetic Fields}
Finally, we allow for $B\ne 0$.   We still demand that $B$ is perturbatively small, so that $\omega_{\mathrm{c}}$ is a perturbative parameter.    As each component of the memory matrix is first order within perturbation theory (we will see this does not change when $B\ne 0$), it will suffice to consider only the $B$-dependent corrections to the memory matrix -- considering $B$-dependent corrections to the static susceptibilities is a higher order correction.       Furthermore, within the memory matrix, any $B$-dependent correction to a parity even coefficient such as $M_{P_xP_x}$ must be $\mathcal{O}(B^2)$, which is second order in perturbation theory.    The only matrix which may admit first order corrections within perturbation theory that are linear in $B$ is the time-reversal non-invariant matrix $N$.

Let us consider the consequences of $B\ne 0$ on the matrix $N$, which relates to static susceptibilities.   The consequences of an external magnetic field are that \begin{equation}
\dot{P}_i = B \epsilon_{ij}J_j^{\mathrm{tot}} + \cdots,   \label{dpb}
\end{equation}with $\mathbf{J}^{\mathrm{tot}}$ the spatially integrated momentum current, where $\cdots$ includes effects such as momentum relaxation, that we have previously accounted for.  
The expression (\ref{dpb}) represents the Lorentz force law, and follows similarly to (\ref{Pdot2}), though with some subtleties.   The (zero momentum component of the) canonical momentum operator $\mathcal{P}_i(\mathbf{x})$ which generates translations is no longer equivalent to the physical momentum:  \begin{equation}
\mathcal{P}_i = P_i + \int \mathrm{d}^2\mathbf{x} \; nA_{B,i}, 
\end{equation}where $\mathbf{A}_B$ is the gauge potential due to the externally imposed magnetic field.   Note that only $P_i$ is gauge invariant.   In addition, the Hamiltonian is altered to $H\rightarrow H+ H_B$, with \begin{equation}
H_B = - \int \mathrm{d}^2\mathbf{x}\; \mathbf{J} \cdot \mathbf{A}_B.
\end{equation}   Now we evaluate $\dot{P}_x$ using the convenient gauge choice $\mathbf{A}_B = -By \hat{\mathbf{x}}$: \begin{align}
\dot{P}_x &= \mathrm{i}[H_0+H_B, P_x ]  = \mathrm{i}\left[H_0+H_B, \mathcal{P}_x + \int \mathrm{d}^2\mathbf{x}\; nBy \right] = \mathrm{i}[H_B, \mathcal{P}_x]  +  \int \mathrm{d}^2\mathbf{x} \; \dot{n}By + \mathcal{O}(B^2) \notag \\
&= -\int \mathrm{d}^2\mathbf{x} \left( -\partial_x J_x - \dot{n}\right) By  = -\int \mathrm{d}^2\mathbf{x} \; \partial_y J_y By = \int \mathrm{d}^2\mathbf{x}\; BJ_y = BJ_y^{\mathrm{tot}}.
\end{align}
As $B$ is a constant, this relates the total current in the $y$ direction to $\dot{P}_x$, as claimed in (\ref{dpb}).   
A similar argument works for $\dot{P}_y$, with the gauge $\mathbf{A}_B = Bx\hat{\mathbf{y}}$.

We can use (\ref{dpb}) to find that the $B$ dependent corrections to the matrix $N$, as $\chi_{\dot{P}_x A} = B \chi_{J_yA}$ for any operator $A$, as an example.   Using this logic, we find  \begin{subequations}\begin{align}
N_{P_x P_y} &= -N_{P_yP_x} = \chi_{P_x\dot{P}_y} = -B\chi_{J_xP_x} = -B\mathcal{Q}, \\
N_{\partial_x n P_y} &= -N_{\partial_y n P_x} = -B\chi_{\partial_y n J_y} = -Bk_x\sigma_{\mathrm{Q}}.
\end{align}\end{subequations}
In writing down these results we have used that $\chi_{P_x\partial_x n} = \chi_{P_y\partial_y n} = 0$, which follows from translation invariance of the thermodynamic state.

To recap:  the ``full memory matrix" $\mathfrak{m}(\omega)$ at this order in perturbation theory is\begin{align}
\mathfrak{m}(\omega) &\equiv M(\omega) + N-\mathrm{i}\omega \chi \approx \left(\begin{array}{cccc} k_x^2\sigma_{\textsc{q}} &\ 0 &\ 0 &\ -Bk_x\sigma_{\textsc{q}} \\ 0 &\ k_x^2\sigma_{\textsc{q}} &\ Bk_x\sigma_{\textsc{q}} &\ 0 \\ 0 &\  -Bk_x\sigma_{\textsc{q}} &\ \mathcal{M}\left(\dfrac{1}{\tau}-\mathrm{i}\omega\right) &\ -B\mathcal{Q} \\  Bk_x\sigma_{\mathrm{Q}}  &\ 0 &\ B\mathcal{Q} &\  \mathcal{M}\left(\dfrac{1}{\tau}-\mathrm{i}\omega\right) \end{array}\right).  \label{mfull}
\end{align}Along with $\chi_{J_xP_x}=  \chi_{J_yP_y} = \mathcal{Q}$,  $\chi_{J_x\partial_x n } = \chi_{J_y\partial_y n} = \sigma_{\textsc{q}}k_x$, being the only non-vanishing static susceptibilities, it is straightforward to invert $\mathfrak{m}$ and obtain the main results 
displayed in (\ref{eq:mainres}), using (\ref{mmf}).

\section{Dynamics of Charge, Heat and Momentum}\label{appa}
Let us now relax the assumption that heat transport has decoupled from the problem, and compute the electrical conductivity using the memory matrix formalism.   In fact, regardless of how many other long lived scalar quantities we have -- governed by diffusive transport -- we will see that (\ref{eq:mainres}) continues to hold.\footnote{Indeed, such conserved quantities cannot at zero momentum have any overlap with the electrical current, as the latter is a vector and we have assumed rotational invariance in this paper.}   In practice, the only additional conserved quantity is heat.   
 
Let us suppose that we have currents $\mathbf{J}_R$ associated with long lived densities $n_R(\mathbf{x})$;  the chemical potentials associated to these long lived densities are $\mu_R$;  the indices $RS\cdots$ will refer to the conserved charges, and indices $ij\cdots$ will refer to spatial indices as usual.    Analogous to (\ref{jmu}) we find \begin{equation}
\mathbf{J}_R = -\Sigma_{RS}\nabla \mu_S,   \label{jsigmars}
\end{equation} with $\Sigma$ a conductivity matrix.  One finds analogously a susceptibility matrix $\chi_{RS}$, and a diffusion matrix $D_{RS}$, with $\mathbf{J}_R = -D_{RS}\nabla n_S$, obeying the Einstein relation \begin{equation}
\Sigma =  D \chi .
\end{equation}
Here we are multiplying together matrices with $RS$ indices.  These results are reviewed in \cite{hartnollscale1}.  The analogous Green's function to $G^{\mathrm{R}}_{nn}$ is \begin{equation}
G^{\mathrm{R}}_{n_Rn_S} = \left(k_x^2 D-\mathrm{i}\omega\right)^{-1}_{RT}k_x^2 \Sigma_{TS},
\end{equation}and so an analogous computation to before gives us \begin{equation}
M_{\partial_x n_R \partial_x n_S} = k_x^2 \Sigma_{RS}
\end{equation}  

Let the index $Q$ denotes the conserved charge;  thus \begin{equation}
\sigma_{\textsc{q}} \equiv \Sigma_{QQ}.
\end{equation}Denote with $\mathbb{P}$ the projection operator onto the $Q$ index: i.e., \begin{equation}
\sigma_{\textsc{q}}\mathbb{P} = \mathbb{P}\Sigma\mathbb{P}.   \label{eqproj}
\end{equation}
$\mathfrak{m}$ generalizes to \begin{equation}
\mathfrak{m} \approx \left(\begin{array}{cccc} k_x^2\Sigma_{RS} &\ 0 &\ 0 &\ -Bk_x\Sigma_{RQ} \\ 0 &\ k_x^2\Sigma_{RS} &\ Bk_x\Sigma_{RQ} &\ 0 \\ 0 &\  -Bk_x\Sigma_{QR} &\ \mathcal{M}\left(\dfrac{1}{\tau}-\mathrm{i}\omega\right) &\ -B\mathcal{Q} \\  Bk_x\Sigma_{QR}  &\ 0 &\ B\mathcal{Q} &\  \mathcal{M}\left(\dfrac{1}{\tau}-\mathrm{i}\omega\right) \end{array}\right).    \label{mmfinal}
\end{equation}
Our goal is to compute \begin{equation}
\sigma_{J_iJ_j} = \chi_{J_i A} \mathfrak{m}^{-1}_{AB}\chi_{BJ_j},   \label{sigmadef2}
\end{equation}and to simply prove that this matrix product is independent of any additional conserved quantities.   (It is straightforward to carry out the remainder of the algebra and explicitly recover the conductivities.)   

A rather generic and useful result for us will be following formula from linear algebra.   Suppose that we have a block matrix \begin{equation}
U = \left(\begin{array}{cc} X &\ Y \\ Z &\ W \end{array}\right),
\end{equation} with $X$ an $m\times m$ matrix, $Y$ an $m\times n$ matrix, $Z$ an $n\times m$ matrix, and $W$ an $n\times n$ matrix.   Then \begin{align}
U^{-1} &= \left(\begin{array}{cc}  (X-YW^{-1}Z)^{-1} &\  -X^{-1}Y(W-ZX^{-1}Y)^{-1} \\ -W^{-1}Z(X-YW^{-1}Z)^{-1} &\ (W-ZX^{-1}Y)^{-1}   \end{array}\right) \notag \\
&=   \left(\begin{array}{cc}  (X-YW^{-1}Z)^{-1} &\  -(X-YW^{-1}Z)^{-1}YW^{-1} \\ -(W-ZX^{-1}Y)^{-1}ZX^{-1} &\ (W-ZX^{-1}Y)^{-1}   \end{array}\right). 
\end{align}

A tedious set of calculations, repeatedly imploying the identities above, leads to the following block matrices in $\mathfrak{m}^{-1}$: \begin{subequations}\label{eq58}\begin{align}
\mathfrak{m}_{P_i P_j}^{-1} &= \mathcal{N} \left( B^2\sigma_{\textsc{q}} + \mathcal{M}(\tau^{-1}-\mathrm{i}\omega)\right)\delta_{ij} + \mathcal{N}B\mathcal{Q}\epsilon_{ij} , \\
\mathfrak{m}_{P_i \partial_j n_R}^{-1} &= \frac{B}{k_x}\delta_{RQ} \mathfrak{m}_{P_i P_k}^{-1} \epsilon_{kj}, \\
\mathfrak{m}_{\partial_i n_R P_j}^{-1} &=  \frac{B}{k_x} \delta_{RQ} \epsilon_{ik} \mathfrak{m}_{P_k P_j}^{-1}, \\
\mathfrak{m}_{\partial_i n_R \partial_j n_S}^{-1} &= \frac{1}{k_x^2}\left[\left(\Sigma^{-1}_{RS} - \mathcal{N}B^2(\mathcal{M}(\tau^{-1}-\mathrm{i}\omega)+B^2\sigma_{\textsc{q}})\mathbb{P}_{RS}\right)\delta_{ij} - \mathcal{N}B^3 \mathcal{Q}\mathbb{P}_{RS}\epsilon_{ij}\right] . \label{eq53d}
\end{align}\end{subequations}
where \begin{equation}
\mathcal{N} = \frac{1}{\mathcal{Q}^2B^2 + \left(B^2\sigma_{\textsc{q}} + \mathcal{M}(\tau^{-1}-\mathrm{i}\omega)\right)^2}.
\end{equation}To obtain (\ref{eq53d}), the following identity, along with (\ref{eqproj}), is helpful (below $c$ is any constant):\begin{equation}
\left(\Sigma + c\Sigma \mathbb{P}\Sigma\right)^{-1} = \Sigma^{-1} - \frac{c}{1+c\sigma_{\textsc{q}}} \mathbb{P}.
\end{equation}
Using (\ref{jsigmars}), we find the electric current - conserved density susceptibilities to be \begin{equation}
\chi_{J_i \partial_j n_R} = \delta_{ij} \Sigma_{QR} k_x.
\end{equation}
It is now a simple matter to see from (\ref{sigmadef2}) that the only component of $\Sigma_{RS}$ that enters the final answer is $\Sigma_{QQ}=\sigma_{\textsc{q}}$;  thus the magnetohydrodynamic result for the electrical conductivity matrix, given in (\ref{eq:mainres}), is unaltered by the presence of additional conserved scalar charges.

\section{Thermal and Thermoelectric Transport}\label{sec:therm}
It is now clear how to extend our work to study thermoelectric transport.   When describing thermoelectric transport, one must specify a set of three matrices:  $\sigma_{ij}$ from before, the Seebeck coefficient(s) $\alpha_{ij}$, and $\bar\kappa_{ij}$, defined by: \begin{equation}
\left(\begin{array}{c} J_i \\ Q_i \end{array}\right) = \left(\begin{array}{cc} \sigma_{ij} &\ \alpha_{ij} \\ T\alpha_{ij} &\ \bar\kappa_{ij} \end{array}\right) \left(\begin{array}{c} E_j \\ -\partial_j T \end{array}\right)
\end{equation} with $Q_i$ the heat current.     We follow the notation of \cite{hkms}, where $\bar\kappa$ is defined as the linear response coefficient between the heat current and the temperature gradient at vanishing electric field;  often in experiments one measures the thermal conductivity $\kappa_{ij}$,  the coefficient between the heat current and temperature gradient at vanishing electric current, $J_i =0$.   The two are related via  \begin{equation}
\kappa = \bar\kappa - T \alpha \sigma^{-1}\alpha.   \label{eq64}
\end{equation}

In the memory matrix framework, we have: \begin{subequations}\begin{align}
\alpha_{ij} T &= \chi_{J_i A} \mathfrak{m}^{-1}_{AB}\chi_{BQ_j}, \\
\bar\kappa_{ij} T &= \chi_{Q_iA} \mathfrak{m}^{-1}_{AB}\chi_{BQ_j}.
\end{align}\end{subequations}
The new susceptibilities we need relate to the heat current.   Letting the index $H$ (in the notation of Section \ref{appa}) denote the diffusive scalar quantity heat, we obtain \begin{equation}
\chi_{Q_i \partial_j n_R} = \delta_{ij}\Sigma_{HR} k_x.
\end{equation}
There is also a new susceptibility: \begin{equation}
\chi_{Q_i P_j} \equiv  \delta_{ij} T\mathcal{S}.
\end{equation}
This serves as a definition of the new quantity, $\mathcal{S}$, which plays a role of ``entropy density".  Using the results of (\ref{eq58}) thermoelectric transport coefficients may now be computed.     Denoting \begin{subequations}\begin{align}
\Sigma_{HQ} &\equiv \alpha_{\textsc{q}} T, \\
\Sigma_{HH} &\equiv \bar\kappa_{\textsc{q}} T,
\end{align}\end{subequations}we obtain \begin{subequations}\label{eq:mainres2}\begin{align}
\alpha_{xx} = \alpha_{yy} &= \frac{(\mathcal{M}(\tau^{-1}-\mathrm{i}\omega)+B^2\sigma_{\textsc{q}})\alpha_{\textsc{q}} + \mathcal{SQ}}{\mathcal{Q}^2B^2 + \left(B^2\sigma_{\textsc{q}} + \mathcal{M}(\tau^{-1}-\mathrm{i}\omega)\right)^2} \mathcal{M}\left(\frac{1}{\tau}-\mathrm{i}\omega\right), \\
\alpha_{xy} = -\alpha_{yx} &= \frac{\alpha_{\textsc{q}}\mathcal{QM}(\tau^{-1}-\mathrm{i}\omega)+\mathcal{S}(\mathcal{Q}^2+B^2\sigma^2_{\textsc{q}}+\sigma_{\textsc{q}}\mathcal{M}(\tau^{-1}-\mathrm{i}\omega))}{\mathcal{Q}^2B^2 + \left(B^2\sigma_{\textsc{q}} + \mathcal{M}(\tau^{-1}-\mathrm{i}\omega)\right)^2}B, \\
\bar\kappa_{xx} = \bar\kappa_{yy} &= \bar\kappa_{\textsc{q}} +  \frac{(B^2\sigma_{\textsc{q}} + \mathcal{M}(\tau^{-1}-\mathrm{i}\omega))(\mathcal{S}^2 - B^2\alpha^2_{\textsc{q}})-2\mathcal{SQ}\alpha_{\textsc{q}}B^2}{\mathcal{Q}^2B^2 + \left(B^2\sigma_{\textsc{q}} + \mathcal{M}(\tau^{-1}-\mathrm{i}\omega)\right)^2}T ,\\
\bar\kappa_{xy} = -\bar\kappa_{yx} &=  \frac{\mathcal{QS}^2-B^2\mathcal{Q}\alpha_{\textsc{q}}^2 + 2\mathcal{S}\alpha_{\textsc{q}}(B^2\sigma_{\textsc{q}}+\mathcal{M}(\tau^{-1}-\mathrm{i}\omega))}{\mathcal{Q}^2B^2 + \left(B^2\sigma_{\textsc{q}} + \mathcal{M}(\tau^{-1}-\mathrm{i}\omega)\right)^2}BT .
\end{align}\end{subequations}
Together with (\ref{eq:mainres}), this forms the main result of this paper, and is the most general framework for magnetotransport  to date, applicable both to relativistic and non-relativistic systems.

If we specialize to a Lorentz-invariant quantum critical system deformed by a chemical potential (for charge) $\mu$, we may compare to the results found in \cite{hkms} using hydrodynamics.   Using (note that we are setting the effective speed of light to be 1) \begin{subequations}\begin{align}
T\alpha_{\textsc{q}} &= -\mu \sigma_{\textsc{q}}, \\
T \bar\kappa_{\textsc{q}} &= \mu^2 \sigma_{\textsc{q}}, \\
T\mathcal{S} &= \mathcal{M} - \mu \mathcal{Q},
\end{align}\end{subequations}
we find agreement with the results of \cite{hkms}. 
On the other hand, if we set $\alpha_{\textsc{q}} = \bar\kappa_{\textsc{q}}=0$, then these results agree with recent holographic calculations performed at $\omega=0$  \cite{holot1, holot2, holot3} -- see also \cite{holot4, holot5}.   This sheds light into the differences between these papers and hydrodynamics, and suggests how one might go about resolving this issue.

It can be helpful to study $\kappa_{ij}$ instead of $\bar\kappa_{ij}$ -- in the limit $B=0$, $\mathcal{Q}\ne0$, $\tau\rightarrow \infty$,  the former is not singular while the latter is (see, e.g., \cite{rosch2}).   $\kappa_{ij}$ can easily be computed using (\ref{eq64}), given our main results (\ref{eq:mainres}) and (\ref{eq:mainres2}):\begin{subequations}\begin{align}
\kappa_{xx}&=\kappa_{yy} =  \bar\kappa_{\textsc{q}}+\frac{\mathcal{Q}^2\mathcal{S}^2\sigma_{\textsc{q}}-2\mathcal{S}\mathcal{Q}^3\alpha_{\textsc{q}}-B^2\mathcal{Q}^2\sigma_{\textsc{q}}\alpha^2_{\textsc{q}} }{\mathcal{Q}^2B^2\sigma_{\textsc{q}}^2 + \left(\mathcal{Q}^2+ \mathcal{M}(\tau^{-1}-\mathrm{i}\omega)\sigma_{\textsc{q}}\right)^2}   T \notag \\
&\;\;\;\;\;\;\;\;\;\;\;\;\;\;\;\;\;\; +\frac{\mathcal{M}(\tau^{-1}-\mathrm{i}\omega)(\sigma_{\textsc{q}}^2\mathcal{S}^2 - 2\mathcal{QS}\alpha_{\textsc{q}}\sigma_{\textsc{q}}-\alpha^2_{\textsc{q}}\mathcal{Q}^2) - \sigma_{\textsc{q}}\alpha^2_{\textsc{q}}\mathcal{M}^2(\tau^{-1}-\mathrm{i}\omega)^2}{\mathcal{Q}^2B^2\sigma_{\textsc{q}}^2 + \left(\mathcal{Q}^2+ \mathcal{M}(\tau^{-1}-\mathrm{i}\omega)\sigma_{\textsc{q}}\right)^2}T \\
\kappa_{xy} &= -\kappa_{yx} = \frac{(\alpha_{\textsc{q}}\mathcal{Q}-\sigma_{\textsc{q}}\mathcal{S})^2}{\mathcal{Q}^2B^2\sigma_{\textsc{q}}^2 + \left(\mathcal{Q}^2+ \mathcal{M}(\tau^{-1}-\mathrm{i}\omega)\sigma_{\textsc{q}}\right)^2}B\mathcal{Q}T 
\end{align}\end{subequations}
Indeed in the $\tau\rightarrow\infty$ limit, $\kappa_{xx}$ is finite (only the first line contributes) so long as $\mathcal{Q}\ne 0$.    Note that if we first set $\mathcal{Q}=0$, then $\kappa_{xx}$ does become singular when $\tau \rightarrow \infty$.   

\section{Conclusions}
\label{sec:conc}

This paper has developed a model of electric, thermal and thermoelectric transport in strange metals which focuses on the influence of a long-lived momentum mode, along with the diffusion of charge and heat.  
Such a long-lived  momentum mode is found in essentially all condensed matter models of non-Fermi liquids, including those obtained from lattice 
models appropriate for the cuprates \cite{raghu,patel}. The influence of an external magnetic field on such a mode is universally
determined by a few thermodynamic susceptibilities: this was established here by the memory matrix formalism, which can be applied to realistic
models of the cuprates.    

Our results are not valid for systems with spontaneously broken global or gauge symmetries, such as superfluids or superconductors.   In this case, the Goldstone modes associated with the broken symmetry must be consistently included within hydrodynamics, so we expect that they must also be included within the memory matrix.   Undertaking such task would be an interesting generalization of the present work.

The resulting $B$ and $T$ dependence of $\sigma_{ij}$, $\alpha_{ij}$ and $\bar\kappa_{ij}$ was then reduced to the $T$ dependence of
the momentum relaxation time $\tau$,  thermodynamic susceptibilities, and diffusive transport coefficients derived via Einstein relations: $\sigma_{\textsc{q}}$, $\alpha_{\textsc{q}}$ and $\bar\kappa_{\textsc{q}}$. An important feature of this approach is that very different physical processes control the values of $\tau$ and  $\sigma_{\textsc{q}}$, $\alpha_{\textsc{q}}$ and $\bar\kappa_{\textsc{q}}$.    Blake and Donos \cite{blakedonos}
argued that reasonable assumptions for the $T$ dependence of $\tau$ and $\sigma_{\textsc{q}}$ (in particular, $\tau \sim 1/T^2$ and $\sigma_{\textsc{q}} \sim 1/T$) lead to an appealing explanation of the data
on the Hall angle on the cuprates \cite{chienexp}. 

The $B$ dependence of $\sigma_{ij}$, $\alpha_{ij}$ and $\bar\kappa_{ij}$ is explicit, as $B$ is a perturbatively small parameter  in this framework.   Their and our discussions have implicitly ignored the $B$ dependence of $\sigma_{\textsc{q}}$, but this could
be important for understanding experimental thermo-electric data \cite{hartnollkarch,khveshchenko}.  

In combination with other recent studies \cite{raghu,patel,DHSS15,patel2}, our results now provide a route to the computation of transport properties
of strange metals using microscopically realistic models. \\
({\em i\/}) For the theory of the onset of spin-density wave order in metals, there is a clear separation of the 
degrees freedom responsible for the two terms in (\ref{eqb0}). The ``lukewarm'' regions of the Fermi surface far from the ``hot spots''
contribute to the second term in (\ref{eqb0}), associated with the slow decay of the momentum mode: a computations of the values of $\mathcal{Q}$, 
$\mathcal{M}$, and $\tau$ in this framework was provided in Ref.~\cite{patel}. In contrast, the ``intrinsic'' quantum critical conductivity, $\sigma_{\textsc{q}}$,
in the first term of (\ref{eqb0}) 
is a property of the particle-hole symmetric hot spot theory; the scaling limit of this theory has $\mathcal{Q}=0$, and so it provides a direct computation 
of $\sigma_{\textsc{q}}$ \cite{patel2}.\\
({\em ii\/}) The same separation between the lukewarm and hot regions of the Fermi surface applies also to the Higgs critical theory of Ref.~\cite{DHSS15}, 
with their respective contributions leading to the two terms in (\ref{eqb0}).\\
({\em iii\/}) For the case of the nematic critical point in 
two-dimensional metals, 
the crucial role of the total momentum mode was discussed in Ref.~\cite{raghu}.  The field theory of this critical point was employed to compute $\mathcal{Q}$, 
$\mathcal{M}$, and $\tau$, with weak disorder providing the source for momentum relaxation. There is not yet a complete understanding of the value of
$\sigma_{\textsc{q}}$ in such models.\\
With these $B=0$ computations of $\mathcal{Q}$, 
$\mathcal{M}$, $\tau$, and $\sigma_{\textsc{q}}$ in hand, then our present analysis shows that the extension to weak $B\neq 0$ in (\ref{eq:mainres}) follows immediately, 
and can be made on quite general grounds.

\addcontentsline{toc}{section}{Acknowledgements}
\section*{Acknowledgements}
We thank Mike Blake, Richard Davison, Blaise Gout\'eraux and Sean Hartnoll for helpful comments on the manuscript.
This research was supported by the NSF under Grant DMR-1360789, the Templeton foundation, and MURI grant W911NF-14-1-0003 from ARO.
Research at Perimeter Institute is supported by the Government of Canada through Industry Canada 
and by the Province of Ontario through the Ministry of Research and Innovation. 

\begin{appendix}
\titleformat{\section}
  {\gdef\sectionlabel{}
   \Large\bfseries\scshape}
  {\gdef\sectionlabel{\thesection. }}{0pt}
  {\begin{tikzpicture}[remember picture,overlay]
	\draw (-0.2, 0) node[right] {\textsf{Appendix \sectionlabel#1}};
	\draw[thick] (0, -0.4) -- (\textwidth, -0.4);
       \end{tikzpicture}
  }
\titlespacing*{\section}{0pt}{15pt}{20pt}

\section{The Memory Matrix Element $M_{P_x\partial_xn}$} \label{appb}
Here we discuss the matrix element $M_{P_x\partial_x n}$, when the operator $\mathcal{O}$ which couples to the translational symmetry breaking field is not charge conjugation symmetric.   From the definition of the memory matrix, \begin{equation}
M_{P_x\partial_x n} = \frac{1}{\pi\mathrm{i}}\int \mathrm{d}\omega \int \mathrm{d}^d\mathbf{q} \; h(\mathbf{q}) \frac{\mathrm{Im}\left(-\mathrm{i}\omega G^{\mathrm{R}}_{\mathcal{O}\partial_x n}(\mathbf{q},k_x,\omega)\right)}{\omega(\omega - z)},
\end{equation}where we have related $\dot{P}_x$ to $\mathcal{O}$.   Let us discuss the general structure of this Green's function.   In fact, noting that $z$ is just above the real axis, we find that \begin{equation}
M_{P_x\partial_x n} = \int \mathrm{d}^d\mathbf{q} \; h(\mathbf{q}) \mathrm{Re}\left(G^{\mathrm{R}}_{\mathcal{O}\partial_xn}(\mathbf{q},k_x,0)\right).
\end{equation} We know that $G^{\mathrm{R}}_{\mathcal{O}\partial_x n} \sim h$, as this Green's function does not obey translation invariance, and thus must be proportional to $h$.   Furthermore, so long as charge is an exactly conserved quantity, and this Green's function is analytic in a neighborhood of $k_x=0$, we conclude that $G^{\mathrm{R}}_{\mathcal{O}\partial_x n} \sim hk_x$.    Thus we conclude that $M_{P_x\partial_x n} \sim h^2k_x$.

Next, let us consider the corrections to the conductivity.   For simplicity, we focus on the case where $B=0$, but similar considerations will hold in the more general case. The full memory matrix $\mathfrak{m}$ is block diagonal, and, considering only $xx$ indices, is \begin{equation}
\mathfrak{m} = \left(\begin{array}{cc} h^2 \mathcal{B}  &\   h^2k_x \mathcal{A} \\ h^2 k_x \mathcal{A} &\  \sigma_{\textsc{q}}k_x^2 \end{array}\right)
\end{equation}
with $\mathcal{A}$ and $\mathcal{B}$ chosen as functions which are not perturbatively small.    We find \begin{equation}
\mathfrak{m}^{-1} = \frac{1}{h^2k_x^2\mathcal{B}\sigma_{\textsc{q}}-k_x^2h^4\mathcal{A}^2} \left(\begin{array}{cc} \sigma_{\textsc{q}}k_x^2 &\   -h^2k_x \mathcal{A} \\ -h^2 k_x \mathcal{A} &\ h^2 \mathcal{B}  \end{array}\right) \approx \left(\begin{array}{cc} 1/h^2\mathcal{B} &\   - \mathcal{A}/\mathcal{B}\sigma_{\textsc{q}}k_x \\ - \mathcal{A}/\mathcal{B}\sigma_{\textsc{q}}k_x &\ 1/\sigma_{\textsc{q}}k_x^2  \end{array}\right)
\end{equation}The correction to the conductivity as given in (\ref{eqb0}) is \begin{equation}
\delta \sigma_{xx} = -\frac{2\mathcal{AQ}}{\mathcal{B}}.
\end{equation}
This is, in our limit, much smaller than either $\sigma_{\textsc{q}}$ or $\mathcal{Q}^2\tau/\mathcal{M}$.   Logically, we should think of this as much smaller than $\sigma_{\textsc{q}}$ since we argued previously that either $\sigma_{\textsc{q}}$ must be anomalously large, or $\mathcal{Q}$ must be very small, in order for $\sigma_{\textsc{q}} \sim \mathcal{Q}^2\tau/\mathcal{M}$ to be possible.   (If $\tau$ is anomalously large and $\sigma_{\textsc{q}}$ is negligible, then we should generically expect $\delta \sigma_{xx}$ to also be negligible).    Thus, it is consistent to ignore these corrections to the memory matrix at leading order.

\end{appendix}

\end{document}